\documentstyle[aps,prl]{revtex}
\begin{document}
\twocolumn[\hsize\textwidth\columnwidth\hsize\csname@twocolumnfalse\endcsname

\title{Glassy Motion of Elastic Manifolds}
\author{Valerii M. Vinokur$^{1}$, M. Cristina Marchetti$^{2}$ and 
Lee-Wen Chen$^{2}$}
\address{$^1$ Material Science Division, Argonne National Laboratory, 
Argonne,
IL 60439}
\address{$^2$ Physics Department, Syracuse University, Syracuse, NY 13244}
\date{\today}

\maketitle

\begin{abstract}
We discuss the low-temperature dynamics of an elastic manifold
driven through a random medium. For driving forces well
below the $T=0$ depinning force, the medium advances via thermally 
activated
hops over the energy barriers separating favorable metastable states.
We show that the distribution of waiting times for these hopping processes
scales as a power-law. This power-law distribution naturally yields
a nonlinear glassy response for the driven medium, 
$v\sim\exp(-{\rm const}\times F^{-\mu})$.

\end{abstract}



\vskip1pc]
\narrowtext

The dynamics of elastic manifolds in random media has received an 
intense attention during the last decade and is in a state of rapid 
current 
development. The reasons for this interest are twofold.
On one hand, elastic manifolds in a
random environment are one of the simplest examples of a
glassy system and yet exhibit a very rich static and dynamical behavior.
Secondly, they can be used as models to study both 
nonlinear collective transport in driven disordered
systems, such as charge density waves, 
polymers, driven interfaces, dislocations in solids, and magnetic 
flux lines in type II superconductors, as well as 
stochastic kinetic 
processes, such as stochastic growth and kinetic roughening 
(see \cite{tora,halp} for a review).

The dynamics of driven elastic manifolds is the result of the 
interplay between 
quenched disorder and the interaction among the many elastic 
degrees of freedom that compose the manifold.
A key physical quantity 
is the
average velocity $v$ of the driven manifold
as a function of the applied force $F$. 
At zero temperature there is 
a depinning transition from a pinned state where $v=0$
to a sliding state at a
critical driving force $F_c$.  A finite temperatures washes out the
transition and the mean velocity
is then finite for all 
driving forces. 
For  low temperatures and driving forces well below the $T=0$
depinning threshold, $F_c$, the dynamics of the driven manifold 
is controlled by thermally activated jumps
of correlated  regions
over the pinning energy barriers
separating  different metastable states.   
In this region the mean velocity is highly nonlinear and has been 
evaluated via a scaling approach \cite{IV}, with the result, 
\begin{equation}
v\simeq \exp[-U(F)/T] \label{speed},
\end{equation}
where $U(F)$  is the optimal energy barrier for creep between favorable
metastable states. 
Under the action of the external drive,
sections 
of the manifold that are initially pinned  move to
a more favorable metastable state, determined by
the condition that the energy gain due to the driving force equals 
the elastic deformation and pinning energy of the medium. For $F<<F_c$
large sections of the manifold hop long distances to find
the next optimal-energy state. This yields a large energy barrier
that diverges algebraically with vanishing driving force,
$U(F)\sim(F_c/F)^{\mu}$, with $\mu$ a characteristic exponent.
The form given in Eq. \ref{speed} was first proposed for
the motion of dislocations in crystals \cite{IV}
and is now widely used to describe the low temperature dynamics
of vortex lines in type-II superconductors \cite{tora}.  

An important recent development has been the confirmation by numerical
studies of an elastic string in $d=2$
of the basic assumption 
of the scaling approach \cite{IV} that the barriers between the 
metastable states scale with the length of the fluctuating string segment
in the same way as the 
fluctuations in the free energy (aside from logarithmic corrections)
\cite{mikh}.
Despite a substantial analytical \cite{nat,nar,makse1} and numerical 
\cite{dong,kaper,lesch1,lesch2,tang,makse2} 
study of the dynamics of  driven disordered elastic 
manifolds,
many open questions remain.

In this Letter we focus on the nature of the 
mechanism by
which the manifold selects 
the appropriate optimal segment controlling the dynamics.
For simplicity  we address this question for the case of
an elastic
string (magnetic flux line or interface) driven through two-dimensional
point disorder.  The paper is organized as follows.  First we 
briefly review the results of the scaling theory of the thermally 
activated motion of elastic manifolds in disordered media following
\cite{IV}. We then develop 
a description of the avalanche-like string motion.

The Hamiltonian of a $D$-dimensional elastic manifold 
driven through a $d$-dimensional disordered 
medium is
\begin{equation}
{\cal H} = \int{d^d}x\left[ \frac{C}2\left( 
\frac{\partial
{{\bf u}}}{\partial {\bf x}}\right) ^2+U\left({\bf x},{\bf u}\right)-{\bf 
F}\cdot{\bf u}
\right],\label{ham}
\end{equation}
where $C$ denotes an elastic stiffness constant (e.g., the linear tension 
of the string; for simplicity we consider the isotropic medium) and ${\bf
u}({\bf x},t)$ is the $n$-dimensional 
transverse displacement field of the manifold.  Eq. (\ref{ham}) 
describes for instance an 
elastic string ($D=1$) in two ($d=2$, $n=1$) and three ($d=3$, $n=2$) 
dimensions, and two- or three-dimensional vortex lattices 
($d=2$, $D=2$, $n=2$ and $d=3$, $D=3$, $n=2$, respectively).
The disorder consists of uncorrelated point defects of density
$n_i$.
It is described by a random 
potential $U\left({\bf x},{\bf u}\right)$ of range $\xi$  and variance
$\Delta=v_0\sqrt{n_i\xi^{d}}$, with $v_0$ the maximum depth of the
potential well of a single pinning center. In the absence of driving force
the string adjusts to the random landscape and traverses the 
medium along rough optimal
paths determined by balancing 
the elastic and pinning energies.
The geometry of these optimal paths is characterized by the 
roughness of the manifold, defined
as 
$w(L)={\langle\left[{\bf u}({\bf x}+{\bf L})-{\bf u}({\bf
x})\right]^2\rangle}^{1/2}$, 
where 
$\langle...\rangle$ denotes the average over both thermal fluctuations and 
quenched
disorder. At large distances the 
roughness
scales as
$w\approx \xi\left(L/L_c\right)^{\zeta}$, where $\zeta<1$ is the roughness 
exponent
and $L_c$ is the pinning correlation length.  
We consider the case of weak 
disorder such that $L_c\gg\xi$. In this regime
domains of linear size $L_c=\xi\left(C\xi^D/\Delta\right)^{2/(4-D)}$
are pinned coherently by disorder when $F=0$.
The pinning length $L_c$  is the smallest scale on which barriers 
between metastable states exist at $F=0$. 
The minimum average energy barrier between neighboring metastable
positions of a pinned segment
$L_c$ is $U_c=C\xi^2/L_c^{(D-2)}$.
Optimal metastable configurations of 
sizes 
$L>L_c$ are then separated by barriers $U(L)\simeq 
U_c(L/L_c)^{2\zeta+D-2}$.
At $T=0$ the string starts to slide when the applied force $F$ can depin
a region of linear size $L_c$, yielding a threshold force
for sliding $F_c=C\xi/L_c^2$.  

We now  consider the dynamics at a finite but low temperature $T\ll U_c$, 
where 
the elementary pinning scales  $U_c$ and $L_c$ are not renormalized 
significantly 
by thermal fluctuations \cite{tora}.  Under the action of the driving 
force a domain of size 
$L$ can be displaced to a new more
favorable metastable state. 
For $F\ll F_c$ the dynamics can be described as the
nucleation of an elementary excitation or nucleus of 
size $L$.
The free energy cost for creating such a nucleus is given by
${\cal 
{F}}[L]=U_c\left(L/L_c\right)^{2\zeta+D-2} -
FL_c^D\xi\left(L/L_c\right)^{\zeta+D}$.
The size $L_{opt}$ of the optimal excitation is obtained by 
minimizing this free energy cost, with the result
$L_{opt}=L_c\left(F_c/F\right)^{1/(2-\zeta)}$.
Nuclei with $L<L_{opt}$ collapse, while 
nuclei of size $L>L_{opt}$ expand. 
In other words excitations on 
scales larger than $L_{opt}$ 
slide freely, while pinning barriers on 
scales $L< L_{opt}$ can be overcome only via thermally 
activated hops. The length scale $L_{opt}$ determines the 
upper bound above which 
thermally activated processes are no longer relevant.  The creep rate is 
determined by the corresponding optimal energy barrier, 
$U(F)={\cal F}(L_{opt})$, with
$U(F)=U_c\left(F_c/F\right)^{\mu}$,
where $\mu=(2\zeta+D-2)/(2-\zeta)$.  For an elastic string in two 
dimensions
$\zeta=2/3$, $D=1$, and $\mu$=1/4.
 
We now show that the form given in Eq. (\ref{speed}) for the mean motion
arises naturally when the driven manifold
sequentially nucleates
all possible excitations on length scales $L<L_{opt}$
to overcome all the possible energy barriers separating 
metastable states on scales $E<U(F)$.  For the sake of simplicity we 
consider
an 
elastic string in $2d$.
We demonstrate 
that the distribution of 
waiting times $\tau(E)$ for hops between metastable states separated by 
energy barriers $E<U(F)$
scales as a {\it power-law},
\begin{equation}
\Psi(\tau)\simeq \tau^{-1-\alpha},\label{wtime}
\end{equation}
with $\alpha<1$.  This
microscopic mechanism of string
dynamics can be viewed as an avalanche-like motion.
The distribution of waiting times given in (\ref{wtime})
is cutoff at the waiting time 
$\tau[U(F)]$ corresponding to the optimal barrier $U(F)$.
For hopping times $\tau$ exceeding $\tau[U(F)]$ only very rare barriers 
with
$E\gg U(F)$ can retard the motion (rare events) and the
distribution function decays much faster than in Eq. (\ref{wtime}).
The average
waiting time that controls the mean string velocity is therefore 
determined by the cutoff
$\tau[U(F)]$.

We begin by considering a segment of finite length $L<L_{opt}$
initially pinned in a 
metastable configuration. We derive an expression for the 
probability $P_L(E)$ that this segment encounters energy barriers smaller
than E when it samples all possible departures from its initial
configuration under the action of the external force $F$.
For any finite driving force the set of all configurations that the 
segment may sample during its activated motion form a connected
cluster of {\it temporarily} pinned states.
To describe this cluster,
we fix the ends of the segment $L$ 
and imagine dividing it in elementary units of longitudinal size $L_c$.
An elementary move consists of the hop of the $i$-th unit
$L_c$ across a transverse distance $\xi$ to a new metastable  
position by overcoming the elementary pinning energy barrier
$U_i$. Here the $\{U_i\}$'s are random variables fluctuating around
their average value, the minimal 
{\it average} pinning barrier, $U_c$. 
The segment
$L$ can be thought of as a singly-connected necklace of $n$ (unit) beads,
where the beads corresponds to the elementary sections $L_c$.
To begin, we consider the simplest class of departures of  $L$
from its initial configuration and describe the advance
of $L$ via the hopping of individual beads over the fluctuating barriers
$U_i$.
The barrier associated with the motion of the segment $L$ is 
then
the largest of the barriers of the elementary hops $L_c$. 
To understand this we note that the hop of a section $L$
of the string to a nearby metastable state can be thought of
as the motion of the domain boundary between spin up and
spin down regions in an Ising spin system with random couplings $\{J_i\}$.
Each elementary unit $L_c$ corresponds to a single Ising spin.
The hop of the $i$-th segment $L_c$ over the corresponding energy
barrier $U_i$ corresponds to a single spin flip.
The flip of spin $i_0$ along the domain boundary will cost an 
energy of order $J_{i_0-1}+J_{i_0+1}$, where $J_{i_0-1}$ and $J_{i_0+1}$
are the couplings to the neighboring spins. The flipping of this
particular spin will facilitate the flipping of the neighboring ones,
since now overturning spin $i_0-1$ will only cost an energy $J_{i_0-2}$,
which
may be larger or smaller than $J_{i_0-1}+J_{i_0+1}$. In either case
the process of overturning the entire domain boundary will be controlled
by the largest energy cost for flipping a single spin.
If $L$ is composed of $n$ elementary segments $L_c$
with energy barriers
$U_i$ randomly fluctuating about $U_c$, then
the energy barrier for the segment $L$ 
is $U=\max{U_i}$. It is important to stress here that we are not trying to
find the optimal hop for the segment $L$ and the associated optimal
energy barrier, but we are simply discussing the distribution of all
possible hops of the segment $L$ from the initial to all 
available final metastable states.  
In order to construct all possible configurations of the advancing segment,
we now imagine redefining the network of elementary hops
by choosing a new  unit $L_1>L_c$ so that the section $L$
is now composed of, say, $n_1$ sub-blocks of size $L_1$
and fluctuating energy barriers $U_{1i}$.
The set of all possible configurations of the excitation $L$ as
it visits all available metastable states forms a new cluster
of elementary excitations or subclusters $L_1$.
Notice that the subclusters of length $L_1$ may in general
be multiply connected, but the argument given above will
still apply upon redefinition of the elementary unit.
The global barrier for the motion of the new cluster 
will again be the maximum of the energy barriers of each subcluster.
The linear structure of the string ensures that the cluster of all possible 
configurations of $L$ remains singly-connected at all levels
of rescaling. It is precisely this linear topology that
enables us to carry out this rescaling  procedure.
Repeated application of this procedure will allow us
to list all possible configurations of the segment in questions and
yield 
the limiting distribution that we seek.
Such a distribution, if it  exists, must
be stable under the $\max$ operation and therefore belongs to the class of
so-called {\it extreme distributions} \cite{gal}.  
To briefly summarize the theory of extreme distributions, we
consider a set of identically distributed independent random
variables $X_i$, with $1\leq i\leq n$ - in the present problem
these are all the energy barriers of the elementary spins and/or block of spins
composing the segment $L$. Let $M_n={\rm max}\{X_1,...,X_n\}$.
If $F(x)$ is the probability of the event $X_i<x$ and ${\cal P}_n(x)$
the probability that $M_n<x$, it can be shown that in the limit
of large $n$ the probability distribution of the maxima can be approximated
by an asymptotic form ${\cal P}(x)$, given by the
solution of the functional equation 
$\lim_{n\rightarrow\infty}{\cal P}_n(x+a_n)
=\lim_{n\rightarrow\infty}F^n(x+a_n)={\cal P}(x)$,
where ${\cal P}(x)$ is defined for $-\infty<x<\infty$.
The extreme distribution we seek 
has the form ${\cal P}(x)=\exp
[-\exp (-x)]$ \cite{gal}.  
The functional form of the distribution must remain invariant 
when the elementary unit of the cluster of all possible hops is 
redefined according to the procedure described earlier.
Since 
$[{\cal P}(x)]^n=\exp[-n\exp(-x)]$,
this requires $a_n=\ln n$.
It thus follows that for the problem of interest here
the appropriate variable $x$ is 
$E-\ln L$, where we measure $E$
in units of $U_c$ and $L$ in units of $L_c$.  The probability 
distribution of
energy barriers for a given segment $L$ is then
${\cal P}_L(E)=\exp [-L\exp (-E)]$. The corresponding distribution density
$p_L(E)=d{\cal P}_L(E)/dE$ is 
\begin{equation}
p_L(E)=Le^{-E}\exp [-L\exp (-E)].\label{distr}
\end{equation}
The typical barrier of a segment of length $L$ scales then as $E\sim U_c\ln
L$.
The global distribution density $W(E)$ of energy barriers for the string is
obtained by integrating $p_L(E)$ over $L$ with the proper weight $n_L$ 
describing 
the density 
of the segments of length $L$. At the critical point $F=0$, the cluster-sizes
distribution is $n_L\sim L^{-\nu}$.
At finite forces $F$ the same  distribution will describe the cluster 
network 
up to $L=L_{opt}$. For the usual percolation 
clusters $\nu =1+d/d_f$, with $d$ the 
dimensionality of the medium and $D<d_f<d$ the fractal 
dimensionality of the clusters network \cite{herrmann}.  For anisotropic 
directed 
percolation describing the depinning transition the exponent may be 
different and will be discussed elsewhere, but the relation $\nu>2$ holds.
By evaluating the integral
we find
\begin{equation}
W(E)\sim e^{-(\nu -1)E/U_c}.\label{W}
\end{equation}
The corresponding probability density of finding a waiting time
$\tau=\tau_\circ \exp (E/T)$, with $\tau_\circ$ a microscopic time scale,
is given by
\begin{equation}
\Psi(\tau)\sim T(\tau_{\circ}/\tau)^{1+\alpha},\,\,\ 
\alpha=(\nu -1)T/U_c.\label{time}
\end{equation}
The distribution of energy barriers given in Eq. \ref{W} must be cut off 
at 
the energy $U(F)$ 
corresponding to the optimal
segment $ L_{opt}(F)$ since the segments of larger scales are, on 
average, sliding freely
and do not participate in the activated dynamics. 
The notion of "waiting time" makes no sense for such excitations.  The 
average
waiting time controlling the motion is therefore
\begin{equation}
\langle\tau\rangle \sim \int^{\tau_{max}} d\tau\Psi(\tau)\tau \sim
\exp[(1-\alpha)U(F)/T], \label{result}
\end{equation}
where $\tau_{max}=\tau_{\circ}\exp[U(F)/T]$. Since the motion is 
controlled by the largest barrier, one arrives at the average velocity 
given by $v\simeq 
u_{opt}/\langle\tau\rangle \simeq \exp[-(1-\alpha)U(F)/T]$, recovering the 
result of the scaling theory \cite{IV}.

We now discuss some observable consequences of the result given in Eq.
\ref{time}.  
The vortex lattice 
in the mixed state of high temperature superconductors is probably
the best system for the experimental study of different aspects of
glassy dynamics, since by tuning parameters
such as the applied field one can probe different regimes.  
For a driven vortex lattice
the  avalanche-like low temperature dynamics just described
will manifest itself in the
spectral density of the density-density 
correlation function,
$S({\bf k}=0,t)=\langle{\delta n_v({\bf r},t)}{\delta n_v({\bf 
r},0)}\rangle$,
which is  accessible by
Hall-probe measurements.
The density-density correlation function can be 
expressed in terms of the correlation function of  the displacement
field $u({\bf r},t)$ as
$S({\bf 
k},t)=k^2\langle{u({\bf k},t)}{u(-{\bf k},0)}\rangle$.  
On large distances displacements 
are uncorrelated in a glassy system and 
$\langle{u({\bf k},t)}{u(-{\bf k},0)}\rangle\propto (1/k^2)f_u({\bf 
k},t)$.  
The 
corresponding spectral density $f_u(\omega)$ can be expressed 
in terms of the 
distribution function of energy barriers $W(E)$,
\begin{equation}
f_u(\omega)\propto\int dEW(E) \frac{\tau (E)}{1+[{\tau (E)}\omega]^2},
\end{equation}
Using $W(E)$ from (\ref{W})  we obtain a power law density noise 
spectrum,
\begin{equation}
S(\omega)\propto \frac{1}{\omega^{1-\alpha}}.
\end{equation}
We see that at low temperatures, $T<U_c$, the exponent $\alpha \ll 1$, and
the noise spectrum of vortex density fluctuations
is nearly the $1/f$ spectrum.

We now turn to discuss the high temperature region, where  
thermal fluctuations renormalize the pinning energy barriers.  
The characteristic 
temperature $T_{dp}$ separating the high temperature and low temperature 
regions is defined by the self-consistent equation $T_{dp}\simeq
U_c(T_{dp})$ 
\cite{tora}.  For elastic 
manifolds with $D\geq 2$, $U_c(T)$ grows much faster than 
$T$ \cite{tora}
for $T>T_{dp}$, and $\alpha<1$.
The results obtained earlier are therefore still relevant, with suitably
renormalized parameters.

The situation is more subtle for the case of an elastic string,
corresponding to $D=1$.  
Above the string depinning temperature, $T_{dp}^s$, the 
minimal pinning energy of the string becomes of order $T$, i.e.,  
$U_c(T)\simeq T$ \cite{IV,tora}. 
The
definition of the depinning temperature used here
follows from a scaling theory based on the postulate that 
the creep barriers scale in the same way 
as the fluctuations in the free energy, i.e., the string statistical 
mechanics 
is controlled by a unique energy scale  \cite{IV,mikh}.
Such a scaling theory leaves numerical 
constants undetermined.  To preserve the
idea that a single energy scale controls the dynamics 
we define $T_{dp}^s$
as the temperature where first  $\alpha(T)=1$.
This definition 
avoids the introduction of a new unphysical
energy 
scale depending on the fractal dimensionality of the cluster of the pinned 
states.  For $T>T_{dp}^s$ the motion 
of the string on scales $L<L_{opt}$ 
is still governed by the waiting times distribution function 
of Eq. \ref{time} with $\alpha=1$, i.e., 
$\Psi(\tau)\sim(1/\tau^2)$.
The average 
waiting time in this high temperature regime is then 
\begin{equation}
\bar \tau_ =\tau_{\circ}\ln(\tau_{max}/\tau_{\circ})\sim 
F^{-\mu}.\label{bartau}
\end{equation}
The characteristic energy and length of the segment hopping with
this waiting time are 
$\bar E\simeq T
\ln(\bar\tau/\tau_{\circ})\sim\ln F$ and 
${\bar L}=L_c(\bar\tau/\tau_{\circ})$, 
respectively.  The typical  transverse displacement ${\bar u}$ of 
a length ${\bar L}$
of string
is determined by $C{\bar u}^2/{\bar L}\simeq F{\bar u}{\bar L}$.
Defining the
string velocity as $v={\bar u}/{\bar \tau}$ and recalling that the activated
velocity contains a prefactor $\sim F$ ensuring that flux flow is recovered
for $F>F_c$, we obtain
\begin{equation}
v\sim (F/F_c)^{2-\mu},\,\,\, T>T_{dp}^s.\label{critglass}
\end{equation}
In the high temperature region, $T>T_{dp}^s$, the string
velocity vanishes as a power law. This represents a kind of marginal glassy
dynamics which may be relevant to the high temperature, low
magnetic field region of type II superconductors. We conjecture that 
this marginal dynamics corresponds to a new glassy state in the
regime of single vortex pinning which can be referred to as {\it critical 
glass}.

In conclusion, we have developed a microscopic model for the glassy 
dynamics 
of elastic manifolds. In our model the low-temperature creep is governed 
by a power law distribution of waiting times.
The dynamics at low temperature, 
$T<T_{dp}$, and long times is dominated by the optimal barrier $U(F)$,
corresponding to the 
maximally pinned configuration. As a result, the mean
velocity of the manifold vanishes exponentially
with vanishing  driving force  according to
$v\simeq \exp[-U(F)/T]$, as obtained 
earlier by a scaling approach \cite{tora}.  
This form for the macroscopic response
arises naturally as a result of the cutoff of the algebraic time 
distribution at the maximal waiting time.  We also conjecture a qualitatively
different marginal glassy response, with $v\propto [(F/F_c(T)]^{1-\mu}$,
for a {\it single} elastic string and
$T>T_{dp}^s$.
Finally, we discuss how the power-law distribution of waiting times
controls the
noise spectrum of the vortex density and can therefore be probed 
experimentally.

This work was supported at Argonne by the U.S. Department of Energy,
BES-Material Sciences, contract No. W-31-109-ENG-38
and at Syracuse by the National Science Foundation, grant No. 
DMR-9419257. VMV acknowledges Stefan Scheidl for illuminating discussions, 
and Pierre LeDoussal for stimulating
conversations in the early stages of this work. L-WC and MCM 
thank Alan Middleton for many helpful discussions.

\end{document}